\documentclass[a4paper]{jpconf}
\usepackage{graphicx}
\begin{document}

\title{Metal-insulator transition and strong-coupling spin liquid in the 
$t{-}t^\prime$ Hubbard model}

\author{Federico Becca, Luca F. Tocchio, and Sandro Sorella}

\address{CNR-INFM Democritos and International School for Advanced Studies 
(SISSA), via Beirut 2-4, 34014 Trieste, Italy}

\ead{becca@sissa.it}

\begin{abstract}
We study the phase diagram of the frustrated $t{-}t^\prime$ Hubbard model on 
the square lattice by using a novel variational wave function.
Taking the clue from the backflow correlations that have been introduced 
long-time ago by Feynman and Cohen and have been used for describing various
interacting systems on the continuum (like liquid $^3$He, the electron
jellium, and metallic Hydrogen), we consider many-body correlations to 
construct a suitable approximation for the ground state of this correlated
model on the lattice. In this way, a very accurate {\it ansatz} can be 
achieved both at weak and strong coupling. We present the evidence that an 
insulating and non-magnetic phase can be stabilized at strong coupling and 
sufficiently large frustrating ratio $t^\prime/t$.
\end{abstract}

The Hubbard model on the square lattice with nearest- and 
next-nearest-neighbor hoppings has been widely studied by many authors with 
different numerical techniques and contradictory outcomes. The Hamiltonian
is given by:
\begin{equation}\label{hubbard}
{\cal H}=-\sum_{i,j,\sigma} t_{ij} c^{\dagger}_{i,\sigma} c_{j,\sigma} + H.c.
+U \sum_{i} n_{i,\uparrow} n_{i,\downarrow},
\end{equation}
where $c^{\dagger}_{i,\sigma} (c_{i,\sigma})$ creates (destroys) an electron
with spin $\sigma$ on site $i$,
$n_{i,\sigma}=c^{\dagger}_{i,\sigma}c_{i,\sigma}$, $t_{ij}$ is the hopping
amplitude (denoted by $t$ and $t^\prime$ for nearest- and next-nearest-neighbor
sites, respectively), and $U$ is the on-site Coulomb repulsion. 
In the following, we will consider the half-filled case, where the number of 
electrons $N$ is equal to the number of sites.
The model of Eq.~(\ref{hubbard}) represents a simple prototype for frustrated 
itinerant materials. In the presence of a finite $t^\prime/t$, there is no 
more a perfect nesting condition that leads to antiferromagnetism for 
any finite $U$ and non-conventional phases may be stabilized at zero 
temperature, like for instance spin liquids with no magnetic order. 

The first numerical study of this model is due to Lin and Hirsch,~\cite{hirsch}
who found the existence of a critical $U_c$ for the appearance of 
antiferromagnetism at finite values of $t^\prime/t$. More recent studies have 
been done by Imada {\it et al.},~\cite{imada1,imada2,imada3} by using the 
Path Integral Renormalization Group approach, 
by Yokoyama {\it et al.},~\cite{ogata} by using a variational Monte Carlo 
method, and by Tremblay {\it et al.},~\cite{tremblay} by a Variational Cluster 
Approximation. Remarkably, all these numerical approaches give very 
different results for the ground-state properties of this simple correlated 
model. In fact, there are huge discrepancies for determining the 
boundaries of various phases, but also for characterizing the most
interesting non-magnetic insulator. Furthermore, also the possibility to have 
superconductivity at small values of $U/t$ is controversial.

In the following, we will show our numerical results, which are based upon
an improved variational Monte Carlo approach that contains backflow 
correlations.~\cite{tocchio} Before doing that, it is useful to remind how
to construct suitable variational states to describe different phases.
Variational wave functions for the unfrustrated Hubbard model,
which has antiferromagnetic long-range order, can be easily constructed by 
considering the ground state $|AF\rangle$ of a mean-field Hamiltonian 
containing a band contribution and a magnetic term
\begin{equation}\label{meanfieldAF}
{\cal H}_{AF}=  -\sum_{i,j,\sigma} 
t_{ij} c^{\dagger}_{i,\sigma} c_{j,\sigma} + H.c. + 
\Delta_{AF} \sum_j e^{i \; {\bf Q} \cdot {\bf R}_j} S_j^x, 
\end{equation}
where $S_j^x$ is the $x$ component of the spin operator
${\bf S}_j=(S_j^x,S_j^y,S_j^z)$ and ${\bf Q}$ is a suitable pitch vector, 
e.g., ${\bf Q}=(\pi,\pi)$ for the Neel phase. In order to have the correct 
spin-spin correlations at large distance, we have to apply a suitable 
long-range spin Jastrow factor, namely 
$|\Psi_{AF}\rangle = {\cal J}_s |AF\rangle$,
with ${\cal J}_s=\exp [-\frac{1}{2} \sum_{i,j} u_{i,j} S_i^z S_j^z ]$,
which governs spin fluctuations orthogonal to the magnetic field
$\Delta_{AF}$.~\cite{becca}
It is important to stress that the mean-field state $|AF\rangle$ can easily 
satisfy the single-occupancy constraint by taking $\Delta_{AF} \to \infty$;
in this limit, it also contains the virtual hopping processes, which are 
generated by the kinetic term, implying that it is possible to reproduce
super-exchange processes.

On the other hand, in pure spin models, namely when $U$ is {\it infinite} and 
charge fluctuations are completely frozen, spin-liquid (i.e., non-magnetic) 
states can be constructed by considering the ground state $|BCS\rangle$ of a 
BCS Hamiltonian with singlet pairing
\begin{equation}\label{meanfieldBCS}
{\cal H}_{BCS}=  -\sum_{i,j,\sigma} 
t_{ij} c^{\dagger}_{i,\sigma} c_{j,\sigma} + H.c. + 
\sum_{i,j} \Delta^{ij}_{BCS} 
(c^{\dagger}_{i,\uparrow} c^{\dagger}_{j,\downarrow} +
c^{\dagger}_{j,\uparrow} c^{\dagger}_{i,\downarrow}) + H.c.
\end{equation}
and then applying to it the so-called Gutzwiller projector,
$|RVB\rangle = {\cal P}_G |BCS\rangle$, where
${\cal P}_G = \prod_i (1-g n_{i,\uparrow} n_{i,\downarrow})$ and
$g=1$.~\cite{anderson} These kind of states can be remarkably accurate 
and represent important tools for the characterization of disordered 
spin-liquid ground states.~\cite{capriotti,yunoki}
However, whenever $U/t$ is finite, the variational state must also contain
charge fluctuations. In this regard, the simplest generalization of the
Gutzwiller projector with $g \ne 1$, which allows doubly occupied sites, is 
known to lead to a metallic phase.~\cite{shiba} One particularly simple way
to obtain a Mott insulator with no magnetic order is to add a sufficiently 
long-range Jastrow factor 
${\cal J}=\exp [-\frac{1}{2} \sum_{i,j} v_{i,j} n_i n_j ]$,
$n_i=\sum_\sigma n_{i,\sigma}$ being the local density.~\cite{capello}
Nevertheless, the accuracy of the resulting wave function
$|\Psi_{BCS}\rangle = {\cal J} |BCS\rangle$ can be rather poor in two 
dimensions for large on-site interactions, especially in presence of 
frustration, since the super-exchange energy scale is not correctly reproduced.
In fact, in contrast to the previous case with magnetic order, within the 
uncorrelated state $|BCS\rangle$ it is not possible to avoid a finite amount 
of double occupancies, and the Gutzwiller factor is mandatory to project out 
high-energy configurations. Here, we propose a simple improvement of (general) 
correlated wave functions in order to mimic the effect of virtual hoppings, 
leading to the super-exchange mechanism. In particular, we want to modify the 
single-particle orbitals, in the same spirit of backflow correlations, 
which have been proposed long time ago by Feynman and Cohen to obtain a 
quantitative description of the roton excitation in liquid 
Helium.~\cite{feynman}
The backflow term has been implemented within quantum Monte Carlo calculations 
to study bulk liquid $^3$He,~\cite{schmidt1,schmidt2} and used to improve the
description of the electron jellium both in two and three
dimensions.~\cite{ceperley1,ceperley2} More recently, it has been also applied
to metallic Hydrogen.~\cite{ceperley3}

Originally, the backflow term corresponds to consider fictitious coordinates 
of the electrons ${\bf r}^b_{\alpha}$, which depend upon the positions of the 
other particles, so to create a return flow of current:
\begin{equation}
{\bf r}^b_{\alpha} = {\bf r}_{\alpha} + \sum_{\beta}
\eta_{\alpha,\beta} [x]  \left ( {\bf r}_{\beta} - {\bf r}_{\alpha} \right ),
\end{equation}
where ${\bf r}_{\alpha}$ are the actual electronic positions and
$\eta_{\alpha,\beta}[x]$ are variational parameters depending in principle
on all the electronic  coordinates, namely on the many-body configuration
$|x\rangle$. The variational wave function is then constructed by means of
the orbitals calculated in the new positions, i.e., $\phi({\bf r}^b_{\alpha})$.
Alternatively, the backflow term can be introduced by considering a linear 
expansion of each single-particle orbital:
\begin{equation}\label{key}
\phi_k({\bf r}^b_{\alpha}) \simeq \phi_k^b({\bf r}_{\alpha}) \equiv
\phi_k({\bf r}_{\alpha}) + \sum_{\beta} c_{\alpha,\beta} [x] \;
\phi_k({\bf r}_{\beta}),
\end{equation}
where $c_{\alpha,\beta}[x]$ are  suitable coefficients that depend on all
electron coordinates.
The definition~(\ref{key}) is particularly useful in lattice models, where the
coordinates of particles may assume only discrete values.
In the Hubbard model, the form of the new ``orbitals'' can be
fixed by considering the $U \gg t$ limit, so to favor a recombination of
neighboring charge fluctuations (i.e., empty and doubly-occupied sites):
\begin{equation}\label{backlattice1}
\phi_k^b({\bf r}_{i,\sigma}) \equiv \epsilon \phi_k({\bf r}_{i,\sigma})
+ \eta \sum_j t_{ij} D_i H_j \phi_k({\bf r}_{j,\sigma}),
\end{equation}
where we used the notation that
$\phi_k({\bf r}_{i,\sigma})= \langle 0|c_{i,\sigma}|\phi_k\rangle$,
$|\phi_k\rangle$ being the eigenstates of the mean-field 
Hamiltonian~(\ref{meanfieldAF}) or~(\ref{meanfieldBCS}),
$D_i=n_{i,\uparrow}n_{i,\downarrow}$, $H_i=h_{i,\uparrow}h_{i,\downarrow}$,
with $h_{i,\sigma}=(1-n_{i,\sigma})$,
so that $D_i$ and $H_i$ are non zero only if the site $i$ is doubly occupied
or empty, respectively; finally $\epsilon$ and $\eta$ are variational
parameters (we can assume that $\epsilon=1$ if $D_iH_j=0$).
As a consequence, already the determinant part of the wave function includes
correlation effects, due to the presence of the many body operator $D_i H_j$.
The previous definition of the backflow term preserves the spin SU(2)
symmetry. A further generalization of the new ``orbitals'' can be made,
by taking all the possible virtual hoppings of the electrons:
\begin{eqnarray}\label{backlattice2}
&&\phi_k^b({\bf r}_{i,\sigma}) \equiv \epsilon \phi_k({\bf r}_{i,\sigma})
+ \eta_1 \sum_j t_{ij} D_i H_j \phi_k({\bf r}_{j,\sigma})
+ \nonumber \\
&& \eta_2 \sum_j t_{ij} n_{i,\sigma} h_{i,-\sigma}
n_{j,-\sigma} h_{j,\sigma} \phi_k({\bf r}_{j,\sigma}) + 
\eta_3 \sum_j t_{ij} \left( D_i n_{j,-\sigma} h_{j,\sigma}
+n_{i,\sigma} h_{i,-\sigma} H_j \right) \phi_k({\bf r}_{j,\sigma}),
\end{eqnarray}
where $\epsilon$, $\eta_1$, $\eta_2$, and $\eta_3$ are variational parameters.
The latter two variational parameters are particularly important for the
metallic phase at small $U/t$, whereas they give only a slight improvement
of the variational wave function in the insulator at strong coupling.
For simplicity, we take the same parameters for up and down electrons.

\begin{figure}
\begin{center}
\includegraphics[scale=0.6]{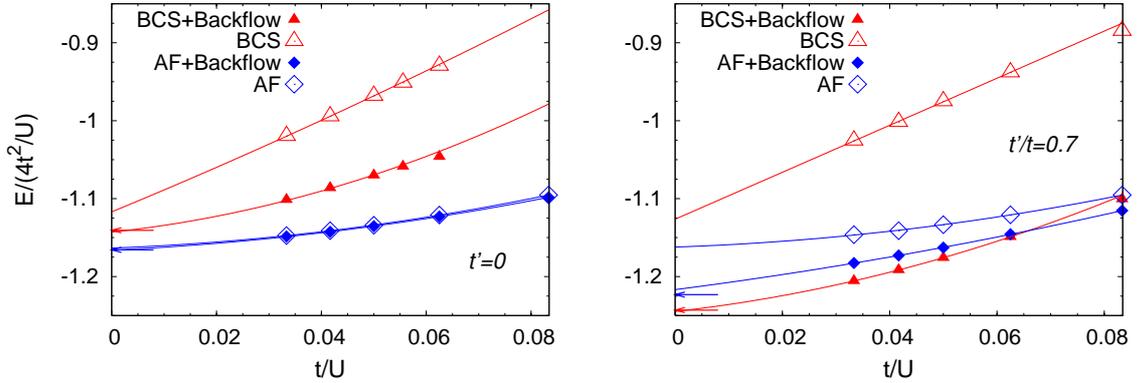}
\caption{\label{estrap} 
Variational energies per site (in unit of $J=4t^{2}/U$) for the BCS state with 
the Jastrow factor, with and without backflow correlations, for 98 sites and
$t^\prime=0$ (left panel) and $t^\prime/t=0.7$ (right panel).
The results for the wave function with antiferromagnetic order and no BCS 
pairing are also shown. Arrows indicate the variational results obtained 
by applying the full Gutzwiller projector (i.e., $g=1$) to the mean-field 
states for the corresponding Heisenberg models.}
\end{center}
\end{figure} 

Thanks to backflow correlations, it is possible to obtain a correct 
extrapolation to the infinite-$U$ limit (i.e., to the variational energy 
obtained with the fully projected states $g=1$ in the Heisenberg model). 
On the contrary, without using backflow terms, the energy of the BCS state, 
even in presence of a fully optimized Jastrow factor, is few hundredths of 
$J=4t^2/U$ higher than the expected value, see Fig.~\ref{estrap}. 
The importance of backflow correlations is even more evident in the frustrated 
case, where they are essential also for improving the accuracy of the 
antiferromagnetic wave function.

\begin{figure}
\begin{center}
\includegraphics[scale=0.6]{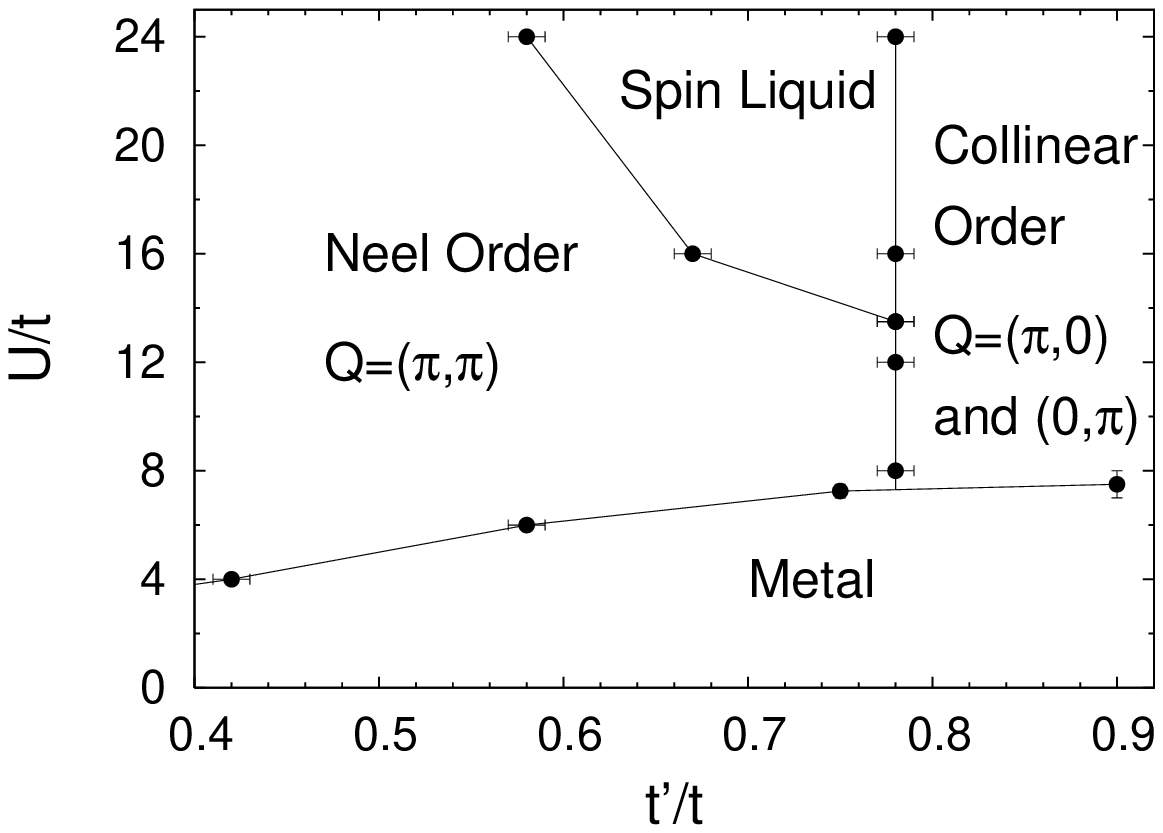}
\includegraphics[scale=0.2]{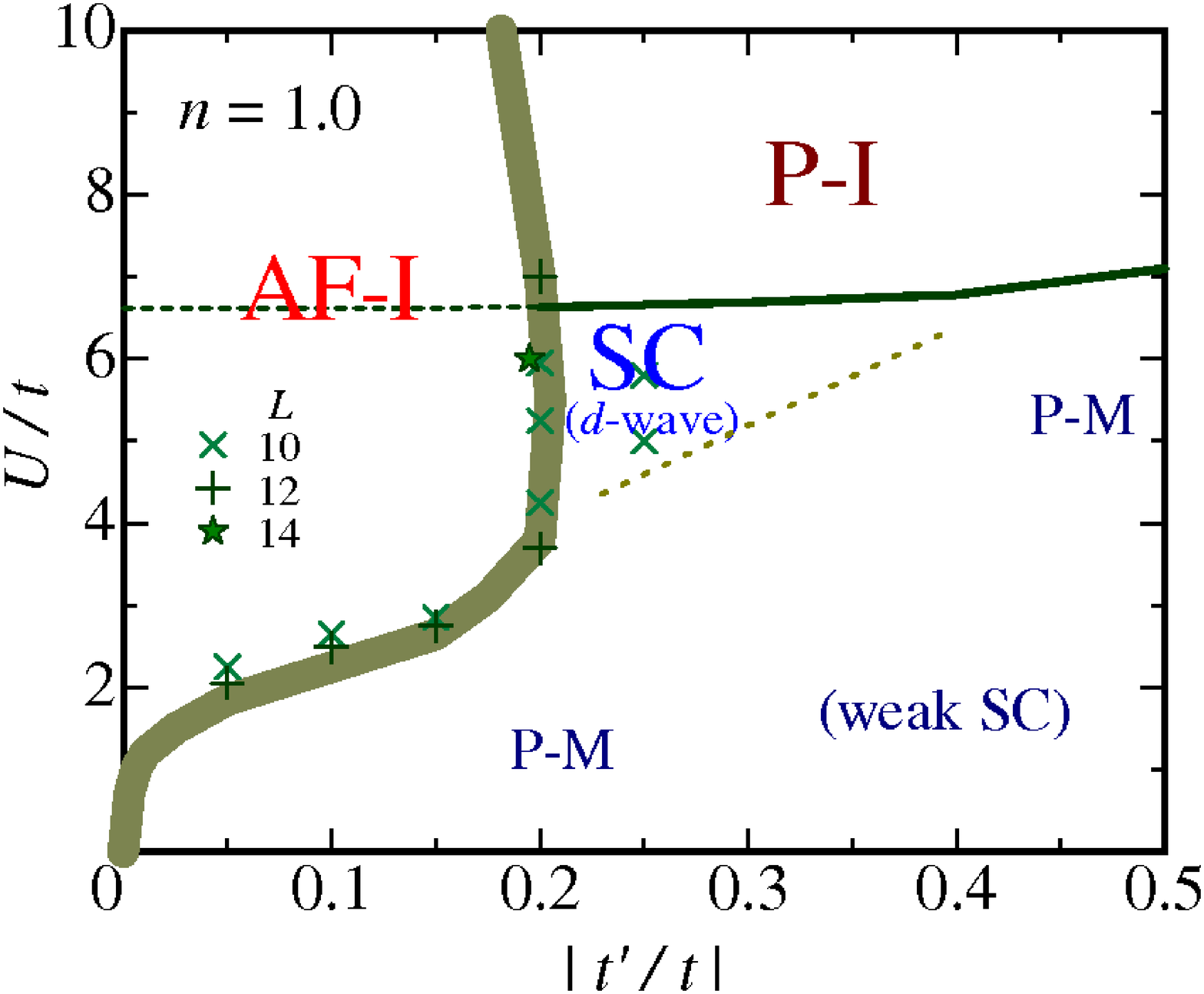}
\caption{\label{myphases} 
Left panel: Phase diagram as obtained by comparing the variational energies of 
different wave functions, all with backflow correlations. Right panel: Phase 
diagram as obtained by Yokoyama {\it et al.} by their variational 
approach.~\cite{ogata}} 
\end{center}
\end{figure}

In order to draw the ground-state phase diagram of the $t{-}t^\prime$ Hubbard 
model, we consider three different wave functions, all with backflow 
correlations: One non-magnetic state $|\Psi_{BCS}\rangle$ and two 
antiferromagnetic states $|\Psi_{AF}\rangle$ with pitch vectors 
${\bf Q}=(\pi,\pi)$ and ${\bf Q}=(\pi,0)$, relevant for small and large 
$t^\prime/t$, respectively. The variational phase diagram is reported in 
Fig.~\ref{myphases}.

\begin{figure}
\includegraphics[scale=0.6]{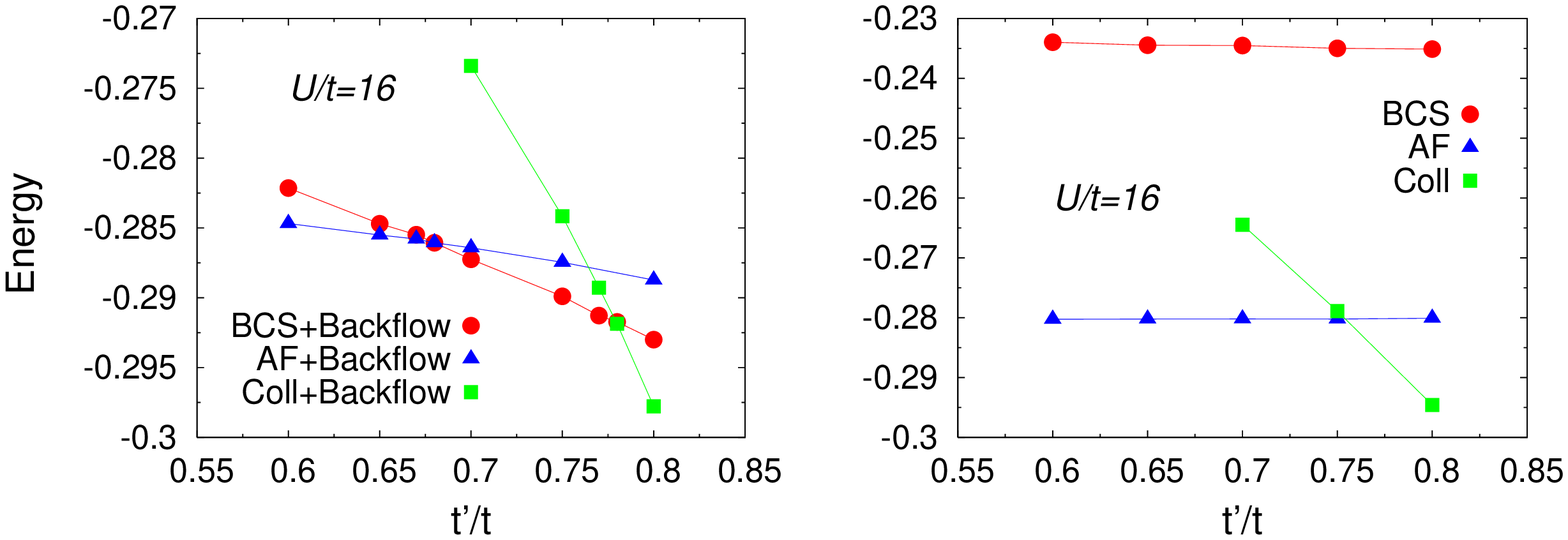}
\caption{\label{energie_U16} 
Variational energies for three different wave functions on the frustrated 
square lattice with and without backflow correlations (left and right panels,
respectively). Red dots denote the energies of the spin-liquid wave function, 
blue triangles the energies of the magnetic state with ${\bf Q}=(\pi,\pi)$, 
and green squares the energies of the magnetic state with ${\bf Q}=(\pi,0)$. 
Data are shown for $U/t=16$ and 98 sites.}
\end{figure} 

The important outcome is that without backflow terms, the energies of the 
spin-liquid wave function are {\it always} higher than those of the 
magnetically ordered states, for any value of the frustration $t^\prime/t$. 
Instead, by inserting backflow correlations, a spin-liquid phase can be 
stabilized at large enough $U/t$ and frustration. For example, this can be 
seen in Fig.~\ref{energie_U16}, where we show the variational energies for 
the  three aforementioned wave functions with and without backflow 
correlations, at $U/t=16$. 

\begin{figure}
\begin{center}
\includegraphics[scale=0.6]{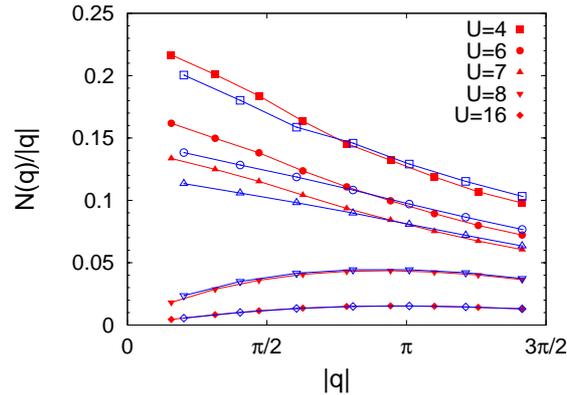}
\caption{\label{Nq} 
Variational results for $N(q)$ divided by $|q|$ for 98 (empty symbols) and 
162 (full symbols) sites and $t^\prime/t=0.75$. 
The metal-insulator transition takes place between $U/t=7$ and $U/t=8$, where 
$N(q)$ changes from a linear to quadratic behavior for $|q|\to 0$}
\end{center}
\end{figure}

In order to study the metal-insulator transition, we look at the static 
density-density correlations $N(q)=\langle n_{-q}n_{q} \rangle$ (where $n_{q}$ 
is the Fourier transform of the local density $n_i$). Indeed, $N(q)$ shows a 
linear behavior for $|q|\to 0$ in the metallic phase and a quadratic behavior 
in the insulating region.~\cite{capello}
For small Coulomb repulsion and finite $t^\prime/t$, $N(q)$ has the linear 
behavior for $|q|\to 0$, typical of a conducting phase. Further, a very small 
superconducting parameter with $d_{x^2-y^2}$ symmetry can be stabilized
(e.g., $\Delta^{ij}_{BCS}=\pm \Delta_{BCS}$ at nearest neighbors)
suggesting that long-range pairing correlations, if any, are tiny. In this 
respect, we compare in Table~\ref{tableI} the optimized $\Delta_{BCS}$ 
when the spin-liquid wave function $|\Psi_{BCS}\rangle$ contains or not 
backflow correlations, for various $U/t$ and $t^\prime/t=0.75$. 
Data show that when accuracy increases, by means of backflow correlations, 
the BCS pairing is reduced by an order of magnitude.
By increasing $U/t$, a metal-insulator transition is found and $N(q)$ acquires 
a quadratic behavior in the insulating phase, indicating a vanishing 
compressibility. In Fig.~\ref{Nq}, we show the variational results for $N(q)$ 
as a function of $U/t$ for $t^\prime/t=0.75$.
The insulator just above the transition is magnetically ordered and the 
variational wave function has a large $\Delta_{AF}$; the transition is likely 
to be first order, since the parameter $\Delta_{AF}$ has a jump across 
the metal-insulator transition. 

In the frustrated regime with $t^\prime/t\sim 0.7$, by further increasing 
$U/t$, there is a second transition to a disordered insulator. Indeed, for 
$U/t > 14$, the energy of the BCS wave function becomes lower than 
the one of the antiferromagnetic state. In this respect, the key ingredient to 
have such an insulating behavior is the presence of a long-range Jastrow 
term ${\cal J}$, which turns a BCS superconductor into a Mott 
insulator.~\cite{ capello} It should be noted that the spin liquid wave 
function contains a superconducting gap with $d_{x^2-y^2}$ symmetry, in 
contrast to what was found in the infinite-$U$ limit, namely in the frustrated 
Heisenberg model, by a similar variational approach.~\cite{capriotti}
In fact, in the latter case, the BCS parameter contains both a term with
$d_{x^2-y^2}$ symmetry and a further term with $d_{xy}$ symmetry.
However, the energy gain due to the latter term is very small in the
Heisenberg model (i.e., order of $0.001J$) and it is very hard to detect it
when charge fluctuations are allowed (i.e., in the Hubbard model).
In all cases that have been analysed, we found that the $d_{xy}$ term is not 
stable in the thermodynamic limit, but converges to zero as the number of 
lattice sites is increased.

\begin{table}
\begin{center}
\begin{tabular}{c|c|c}
$U/t$ & $\Delta_{BCS}$ with backflow & $\Delta_{BCS}$ without backflow\\
\hline
7 & 0.042(1) & 0.306(1) \\
6 & 0.031(1) & 0.145(1) \\ 
4 & 0.012(1) & 0.039(1) \\
2 & 0.002(1) & 0.021(1) \\
\end{tabular}
\caption{\label{tableI} 
BCS pairing $\Delta_{BCS}$ for various $U/t$ in the metallic region at 
$t^\prime/t=0.75$ and 98 sites.}
\end{center} 
\end{table}

\begin{table} 
\begin{center}
\begin{tabular}{c|c|c}
Wave function & Energy ($U/t=20, t^\prime/t=0.7$) & Energy ($U/t=8, t^\prime/t=0.3$) \\
\hline
${\cal J}|BCS\rangle$                     & -0.1950(1)  & -0.4016(1) \\
${\cal J}_{HD}{\cal J}|BCS\rangle$        & -0.2061(1)  & -0.4180(1) \\ 
${\cal J}|BCS + \rm{Backflow}\rangle$     & -0.23516(4) & -0.4879(1) \\
${\cal J}_s|AF + \rm{Backflow}\rangle$    & -0.23257(3) & -0.5222(1)
\end{tabular}
\caption{\label{tableII} 
Variational energies (in unit of $t$) for three spin-liquid wave functions 
and for the best antiferromagnetic state with Neel order. 
The cluster contains 98 sites.  ${\cal J}_{HD}$ is a short-range many-body 
Jastrow factor that has been used in Ref.~\cite{ogata}.}
\end{center}
\end{table}

We can make a direct comparison of our energies with the ones obtained 
by Yokoyama and collaborators,~\cite{ogata} who used a similar variational 
wave function containing a particular many-body Jastrow factor ${\cal J}_{HD}$
to correlate empty and doubly occupied sites at nearest-neighbor distances. 
In Table~\ref{tableII}, we report the variational energy of the simple 
spin-liquid state $|\Psi_{BCS}\rangle ={\cal J}|BCS\rangle$, together with 
the improved energies, which are obtained by adding the Jastrow term 
${\cal J}_{HD}$ or by considering backflow correlations. We notice that the 
latter state always gives much lower energies than the one obtained with the 
additional Jastrow factor. 
In particular, let us consider the case of $U/t=8$ and $t^\prime/t=0.3$, 
which should be magnetically ordered according to our calculations and 
disordered according to Ref.~\cite{ogata} (see Fig.~\ref{myphases}). 
In this case, even though our spin-liquid state has a much better energy than 
the one with ${\cal J}_{HD}$, the best wave function has Neel order, 
indicating that the stability region of antiferromagnetism is larger than what 
predicted by Yokoyama and collaborators.~\cite{ogata} 

To conclude, we compare the variational energies with the ones obtained by
the Green's function Monte Carlo approach, implemented within the Fixed Node
(FN) approximation. In brief, the FN method allows one to filter out the 
high-energy components of a given state and to find the best variational state 
with the same nodes of the starting one.~\cite{ceperleyfn} On the lattice, the 
FN method can be defined as follows: Starting from the original 
Hamiltonian ${\cal H}$, we define an effective Hamiltonian by adding a 
perturbation $O$:
\begin{equation}\label{fnham}
{\cal H}^{eff} = {\cal H} + O.
\end{equation}
The operator $O$ is defined through its matrix elements and depends upon
a given guiding function $|\Psi \rangle$, that is for instance the variational
state itself
\begin{equation}
O_{x^\prime,x} = \left \{
\begin{array}{ll}
-{\cal H}_{x^\prime,x} & {\rm if} \; s_{x^\prime,x} = \Psi_{x^\prime} {\cal H}_{
x^\prime,x} \Psi_x >0 \nonumber \\
\sum_{y,s_{y,x}>0} {\cal H}_{y,x} \frac{\Psi_y}{\Psi_x} & {\rm for} \; x^\prime=
x,
\end{array}
\right .
\end{equation}
where $\Psi_x = \langle x|\Psi \rangle$, $|x\rangle$ being a generic many-body
configuration.
The most important property of this effective Hamiltonian is that its ground 
state $|\Psi_0 \rangle$ can be efficiently computed by using the Green's
function Monte Carlo technique,~\cite{nandini,calandra} which allows one to
sample the distribution
$\Pi_x \propto \langle x|\Psi \rangle \langle x|\Psi_0 \rangle$
by means of a statistical implementation of the power method: 
$\Pi \propto \lim_{n \to \infty } G^n \Pi^0$, where
$\Pi^0$ is a starting distribution and $G_{x^\prime,x}= \Psi_{x^\prime}
(\Lambda \delta_{x^\prime,x} - {\cal H}^{eff}_{x^\prime,x})/ \Psi_x$
is the so-called Green's function, defined with a large or even
infinite positive constant $\Lambda$, $\delta_{x^\prime,x}$
being the Kronecker symbol. The statistical method is very efficient
since in this case  all the matrix elements of $G$ are non-negative and, 
therefore, it can represent a transition probability in configuration space, 
apart for a normalization factor $b_x= \sum_{x^\prime} G_{x^\prime,x}$.
In this case, it follows immediately that the asymptotic distribution $\Pi$
is also positive and, therefore, we arrive at the important conclusion that
the ground state of ${\cal H}^{eff}$ has the same signs of the chosen guiding 
function (i.e., the best variational state).

\begin{figure}
\begin{center}
\includegraphics[scale=0.6]{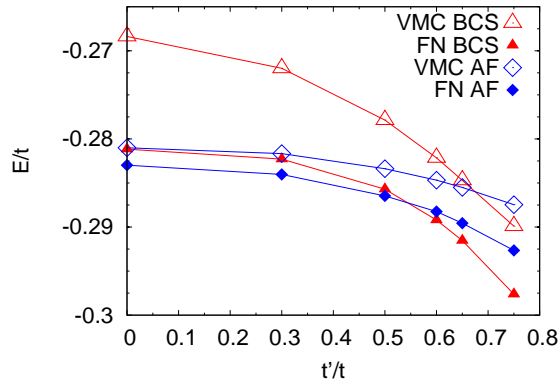}
\caption{\label{GFMCenergies} 
Comparison between the variational (VMC) energies per site (with backflow 
correlations) and the FN ones. Data are shown for $U/t=16$ and 98 sites.}
\end{center}
\end{figure}

In Fig.~\ref{GFMCenergies}, we show the variational energies per site (with 
backflow correlations) and the FN ones for $U/t=16$ on a 98-site lattice. 
The small energy difference between the pure variational energies and the FN 
ones demonstrates the accuracy of the backflow states. Notice that 
$|\Psi_{AF}\rangle$ and $|\Psi_{BCS}\rangle$ have different nodal surfaces, 
implying different FN energies.

\begin{figure}
\begin{center}
\includegraphics[scale=0.6]{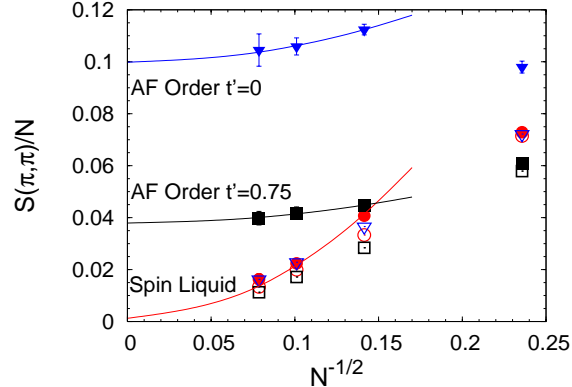}
\caption{\label{GFMCspinspin} 
Variational (empty symbols) and FN (full symbols) results for 
$S(\pi,\pi)/N$, for $N=18$, $50$, $98$, and $162$. All the calculations have 
been done by using the projected BCS wave function; 
$U/t=16$ and $t'/t=0$ (triangles), $U/t=24$ and $t'/t=0.7$ (circles) and 
$U/t=8$ and $t'/t=0.75$ (squares). Lines are guides to the eye.}
\end{center}
\end{figure}

In order to verify the magnetic properties obtained within the variational 
approach, we can consider the static spin-spin correlations 
$S(q)=\langle S^z_qS^z_{-q}\rangle$, where $S^z_q$ is the Fourier transform
of the local spin $S^z_i$. Although the FN approach may break the 
SU(2) spin symmetry, favoring a spin alignment along the $z$ axis, $S(q)$ 
is particularly simple to evaluate within this approach,~\cite{ceperleyfn} and 
it gives important insights into the magnetic properties of the ground state. 
In Fig.~\ref{GFMCspinspin}, we report the comparison between the variational 
and the FN results by using the non-magnetic state $|\Psi_{BCS}\rangle$. 
Remarkably, in the unfrustrated case, where antiferromagnetic order takes
place, the FN approach is able to increase spin-spin correlations at 
${\bf Q}=(\pi,\pi)$, even by considering the non-magnetic wave function to 
fix the nodes. 
In this case, the FN results are qualitatively different from the pure
variational ones, which indicate no magnetic order in the thermodynamic limit.
A finite value of the magnetization is also plausible in the insulating region 
just above the metallic phase at strong frustration 
(i.e., $t^\prime/t\sim 0.75$), confirming our variational calculations. 
On the contrary, by increasing electron correlation, FN results 
change only slightly the variational value of $S(\pi,\pi)$, indicating the 
stability of the disordered state. Therefore, the FN results confirm that a 
spin liquid region can be stabilized only at large enough $U/t$, while the 
insulator close to the metallic region is magnetically ordered.

In summary, the backflow wave functions represent simple and useful
generalizations of standard projected states, which are used to describe
strongly correlated materials. They are highly accurate and may give important 
insights into the ground-state properties of frustrated models with itinerant 
electrons.

\ack We acknowledge CNR-INFM for partial support.


\begin{thebibliography}{9}

\bibitem{hirsch} H.Q. Lin and J.E. Hirsch, Phys. Rev. B {\bf 35}, 3359 (1987).
\bibitem{imada1} T. Kashima and M. Imada, J. Phys. Soc. Jpn. {\bf 70}, 3052
   (2001).
\bibitem{imada2} H. Morita, S. Watanabe, and M. Imada, J. Phys. Soc. Jpn.
   {\bf 71}, 2109 (2002).
\bibitem{imada3} T. Mizusaki and M. Imada, Phys. Rev. B {\bf 74}, 014421 
   (2006).
\bibitem{ogata} H. Yokoyama, M. Ogata, and Y. Tanaka, J. Phys. Soc. Jpn.
   {\bf 75}, 114706 (2006).
\bibitem{tremblay} A.H. Nevidomskyy, C. Scheiber, D. Senechal,
   and A.-M.S. Tremblay, Phys. Rev. B {\bf 77}, 064427 (2008); see also,
   S.R. Hassan, B. Davoudi, B. Kyung, and A.-M.S. Tremblay, Phys. Rev. B 
   {\bf 77}, 094501 (2008).
\bibitem{tocchio} L.F. Tocchio, F. Becca, A. Parola, and S. Sorella,
   Phys. Rev. B {\bf 78}, 041101(R) (2008).
\bibitem{becca} F. Becca, M. Capone, and S. Sorella, Phys. Rev. B {\bf 62}, 
   12700 (2000).
\bibitem{anderson} P.W. Anderson, Science {\bf 235}, 1196 (1987).
\bibitem{capriotti} L. Capriotti, F. Becca, A. Parola, and S. Sorella,
   Phys. Rev. Lett. {\bf 87}, 097201 (2001).
\bibitem{yunoki} S. Yunoki and S. Sorella, Phys. Rev. B {\bf 74}, 014408 (2006).
\bibitem{shiba} H. Yokoyama and H. Shiba, J. Phys. Soc. Jpn. {\bf 56},
  1490 (1987).
\bibitem{capello} M. Capello, F. Becca, M. Fabrizio, S. Sorella,
   and E. Tosatti, Phys. Rev. Lett. {\bf 94}, 026406 (2005).
\bibitem{feynman} R.P. Feynman and M. Cohen, Phys. Rev. {\bf 102}, 1189 (1956).
\bibitem{schmidt1} M.A. Lee, K.E. Schmidt, M.H. Kalos, and G.V. Chester,
   Phys. Rev. Lett. {\bf 46}, 728 (1981).
\bibitem{schmidt2} K.E. Schmidt, M.A. Lee, M.H. Kalos, and G.V. Chester,
   Phys. Rev. Lett. {\bf 47}, 807 (1981).
\bibitem{ceperley1} Y. Kwon, D.M. Ceperley, and R.M. Martin, Phys. Rev. B 
   {\bf 48}, 12037 (1993).
\bibitem{ceperley2} Y. Kwon, D.M. Ceperley, and R.M. Martin, Phys. Rev. B 
   {\bf 58}, 6800 (1998).
\bibitem{ceperley3} M. Holzmann, D.M. Ceperley, C. Pierleoni, and K. Esler,
   Phys. Rev. E {\bf 68}, 046707 (2003).
\bibitem{ceperleyfn} D.F.B. ten Haaf, H.J.M. van Bemmel, J.M.J. van Leeuwen,
   W. van Saarloos, and D.M. Ceperley, Phys. Rev. B {\bf 51}, 13039 (1995).
\bibitem{nandini} N. Trivedi and D.M. Ceperley, Phys. Rev. B {\bf 41}, 4552 
   (1990).
\bibitem{calandra} M. Calandra and S. Sorella, Phys. Rev. B {\bf 57}, 11446 
   (1998).

\end{thebibliography}
\end{document}